\documentclass[prl,twocolumn,superscriptaddress,showpacs]{revtex4}
\usepackage{times}
\usepackage{amsmath}
\usepackage{amsthm}
\usepackage{amssymb}
\usepackage{amsbsy}
\usepackage{graphicx}
\usepackage{pstricks}
\usepackage{color}

\begin{document}
\newcommand{\hexaa}{\;\pspicture(0,0.1)(0.35,0.6)\psset{unit=0.75cm}
\pspolygon(0,0.15)(0,0.45)(0.2598,0.6)(0.5196,0.45)(0.5196,0.15)(0.2598,0)
\psset{linewidth=0.08,linestyle=solid}
\psline(0,0.15)(0,0.45)
\psline(0.2598,0.6)(0.5196,0.45)
\psline(0.5196,0.15)(0.2598,0)
\psdots[linecolor=gray,dotsize=.20](0,0.30)
\psdots[dotstyle=triangle*,linecolor=gray,dotsize=.20](0.3897,0.525)
\psdots[linecolor=gray,dotsize=.20,dotstyle=square*](0.3897,0.075)
\endpspicture\;}

\newcommand{\hexab}{\;
\pspicture(0,0.1)(0.35,0.6)
\psset{unit=0.75cm}
\pspolygon(0,0.15)(0,0.45)(0.2598,0.6)(0.5196,0.45)(0.5196,0.15)(0.2598,0)
\psset{linewidth=0.08,linestyle=solid}
\psline(0,0.15)(0,0.45)
\psline(0.2598,0.6)(0.5196,0.45)
\psline(0.5196,0.15)(0.2598,0)
\psdots[dotstyle=square*,linecolor=gray,dotsize=.20](0,0.30)
\psdots[linecolor=gray,dotsize=.20](0.3897,0.525)
\psdots[dotstyle=triangle*,linecolor=gray,dotsize=.20](0.3897,0.075)
\endpspicture\;}

\newcommand{\hexb}{\;
\pspicture(0,0.1)(0.35,0.6)
\psset{unit=0.75cm}
\psset {linewidth=0.03,linestyle=solid}
\pspolygon[](0,0.15)(0,0.45)(0.2598,0.6)(0.5196,0.45)(0.5196,0.15)(0.2598,0)
\psset{linewidth=0.08,linestyle=solid}
\psline(0.2598,0.6)(0,0.45)
\psline(0.5196,0.15)(0.5196,0.45)
\psline(0.2598,0)(0,0.15)
\psdots[linecolor=gray,dotsize=.20](0.1299,0.525)
\psdots[dotstyle=triangle*,linecolor=gray,dotsize=.20](0.525,0.3)
\psdots[linecolor=gray,dotsize=.20,dotstyle=square*](0.1299,0.075)
\endpspicture\;}

\newcommand{\smallhexh}{ \;
\pspicture(0,0.1)(0.2,0.3)
\psset{linewidth=0.03,linestyle=solid}
\pspolygon[](0,0.0775)(0,0.225)(0.124,0.3)(0.255,0.225)(0.255,0.0775)(0.124,0)
\endpspicture
\;}

\begin{abstract}
We study strongly correlated electrons on a kagom\'{e} lattice at
1/6 (and 5/6) filling. They are described by an extended Hubbard
Hamiltonian. We are concerned with the limit $|t|\ll V\ll U$ with
hopping amplitude $t$, nearest-neighbor repulsion $V$ and on-site
repulsion $U$. We derive an effective Hamiltonian and show, with the
help of the Perron--Frobenius theorem, that the system is
ferromagnetic at low temperatures. The robustness of ferromagnetism
is discussed and extensions to other lattices are indicated.
\end{abstract}

\title{Kinetic ferromagnetism on a kagom\'{e} lattice}

\author{F. Pollmann}
\address{Max-Planck-Institut f{\"u}r Physik komplexer Systeme, 01187 Dresden, Germany}
\author{P. Fulde}
\address{Max-Planck-Institut f{\"u}r Physik komplexer Systeme, 01187 Dresden, Germany}
\address{Asia Pacific Center for Theoretical Physics, Pohan, Korea}

\author{K. Shtengel}

\address{Department of Physics and Astronomy, University of California,
Riverside, CA 92521 }

\maketitle

\paragraph*{Introduction.}

Ferromagnetism in solids or molecules can be of different origin.
The most common is spin exchange between electrons belonging to neighboring
sites. Polarization of the spins and formation of a symmetric spin
state reduces the effects of mutual Coulomb repulsions of the electrons due
to the Pauli exclusion principle. The physics is the same
as for intra-atomic Hund's rule coupling, which also plays a significant
role in the theory of ferromagnetism. The key ingredient of this
mechanism is the \emph{potential} energy of repulsive electron-electron
interactions minimized by a symmetric spin state which is better at
keeping electrons apart.
This should be contrasted with the standard
superexchange mechanism for antiferromagnetism where the \emph{kinetic}
energy of electrons is optimized instead.
Hence it is often the competition between the potential and kinetic energies
that determines the ``winner''. This physics is illustrated, in its extreme
limit, in the case of flat-band ferromagnetism
\cite{Mielke91a,Mielke92}. Mielke pointed out that electrons in a
half-filled flat band become fully spin-polarized for \emph{any} strength of
the on-site repulsion $U$.
(One could also think of
this effect as an extreme case of the Stoner instability in metals.)

Therefore it might  appear surprising that
ferromagnetism can also originate from purely kinetic effects.
A prominent example is the ferromagnetic ground state (GS) discussed by
Nagaoka \cite{Nagaoka66} which is due to the motion of a single hole in an
otherwise half-filled Hubbard system. The argument based on the application of the Perron--Frobenius theorem shall
be presented later. Although it is only valid in the limit of the infinite
on-site Hubbard repulsion (to exclude the possibility of double occupancy) on
a finite lattice, it demonstrates  how ferromagnetism can result from
the motion of the electrons or holes.
The same theorem is also the basis of ferromagnetism due to
three-particle ring exchange, a process first pointed out by Thouless
\cite{Thouless65} in the context of $^{3}$He (following the original
observation by Herring \cite{Herring62}) and later also studied in the
context of Wigner glass \cite{Chakravarty99} and frustrated magnets
\cite{Misguich99}.
In both cases, the ferromagnetic GS has the smoothest
wavefunction and hence lowest kinetic energy.

Our introduction would not be complete without mentioning some
other sources of ferromagnetism such as the RKKY interaction
in metals or double-exchange (e.g., in manganites) to name a few.
In this paper, however, we will be concerned with ferromagnetism of kinetic
origin. In particular, we demonstrate that fermions on a partially
filled kagom\'{e} lattice which are described by an extended one-band
Hubbard model in the strong correlation limit have a ferromagnetic
GS. (The difference with Mielke's flat-band ferromagnetism is
discussed later in the paper.) Again, the physics discussed
 here is motivated by the Perron--Frobenius theorem, but otherwise is quite
different from Nagaoka's and Thouless' examples.

\paragraph*{Model Hamiltonian.}

We start from an extended one-band Hubbard model on a kagom\'{e} lattice
with on-site repulsion $U$ and nearest-neighbor repulsion $V$.
Using second quantized notation, the Hamiltonian is written as
\begin{equation}
H
=-t\sum_{\langle
i,j\rangle,\sigma}\left(c_{i\sigma}^{\dag}c^{\vphantom{\dag}}_{j\sigma}
+ \text{H.c.}\right) \\
  +V\sum_{\langle
i,j\rangle}n_{i}n_{j}+U\sum_{i}n_{i\uparrow}n_{i\downarrow}.
\label{eq:extended_hub}
\end{equation}
Here the operators $c^{\vphantom{\dag}}_{i\sigma}$ ($c_{i\sigma}^{\dag}$)
annihilate (create) an electron with spin $\sigma$ on site $i$. The density
operators are given by $n_{i}=n_{i\uparrow}+n_{i\downarrow}$ with
$n_{i\sigma}=c_{i\sigma}^{\dag}c^{\vphantom{\dag}}_{i\sigma}$. The notation
$\langle i,j\rangle$ refers to pairs of nearest neighbors.

\begin{figure}
\begin{centering}
\begin{tabular}{ccc}
~&
~&
\tabularnewline
(a)\includegraphics[width=28mm]{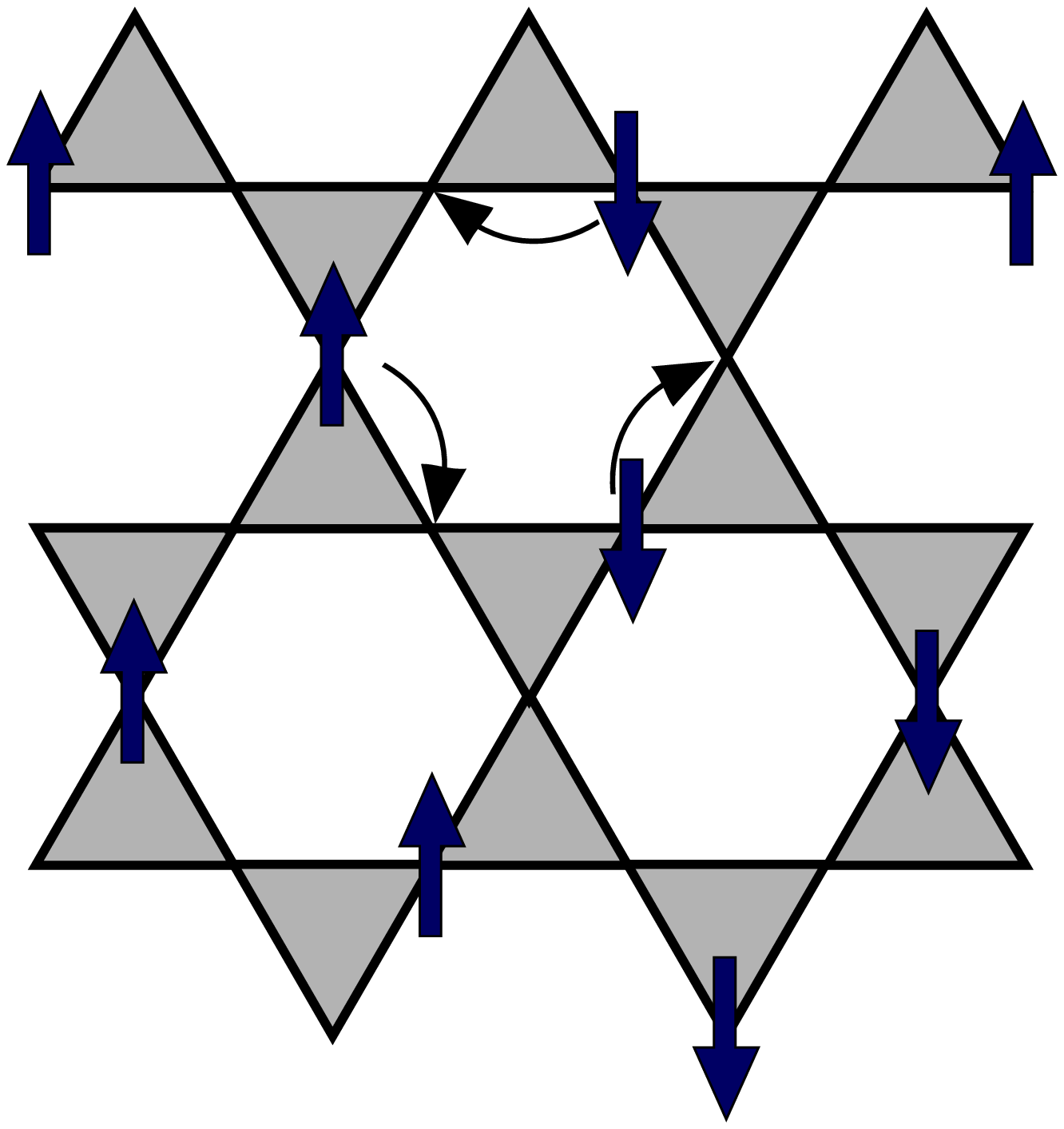}&
(b)\includegraphics[width=28mm]{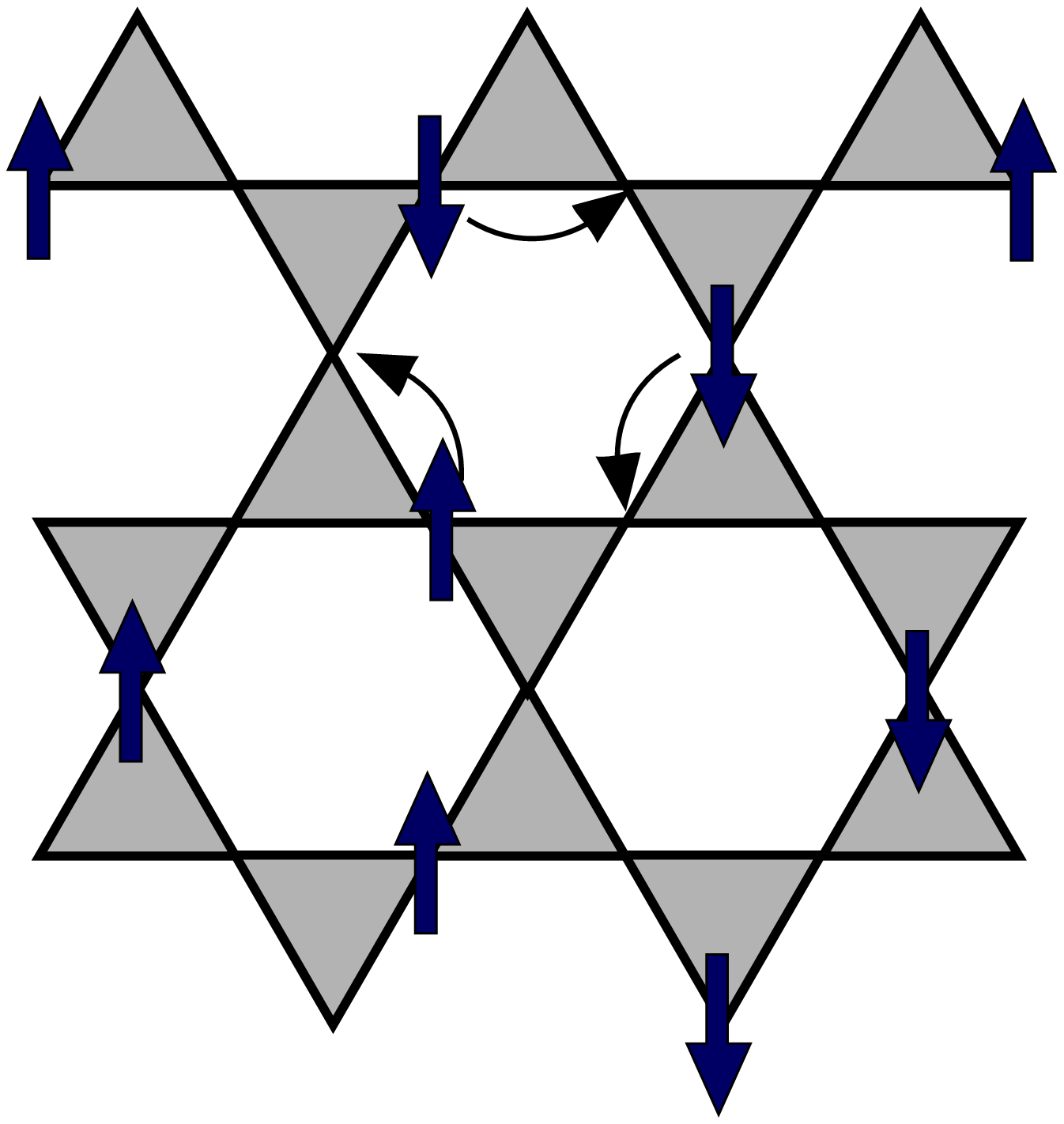}&
\tabularnewline
(c)\includegraphics[width=28mm]{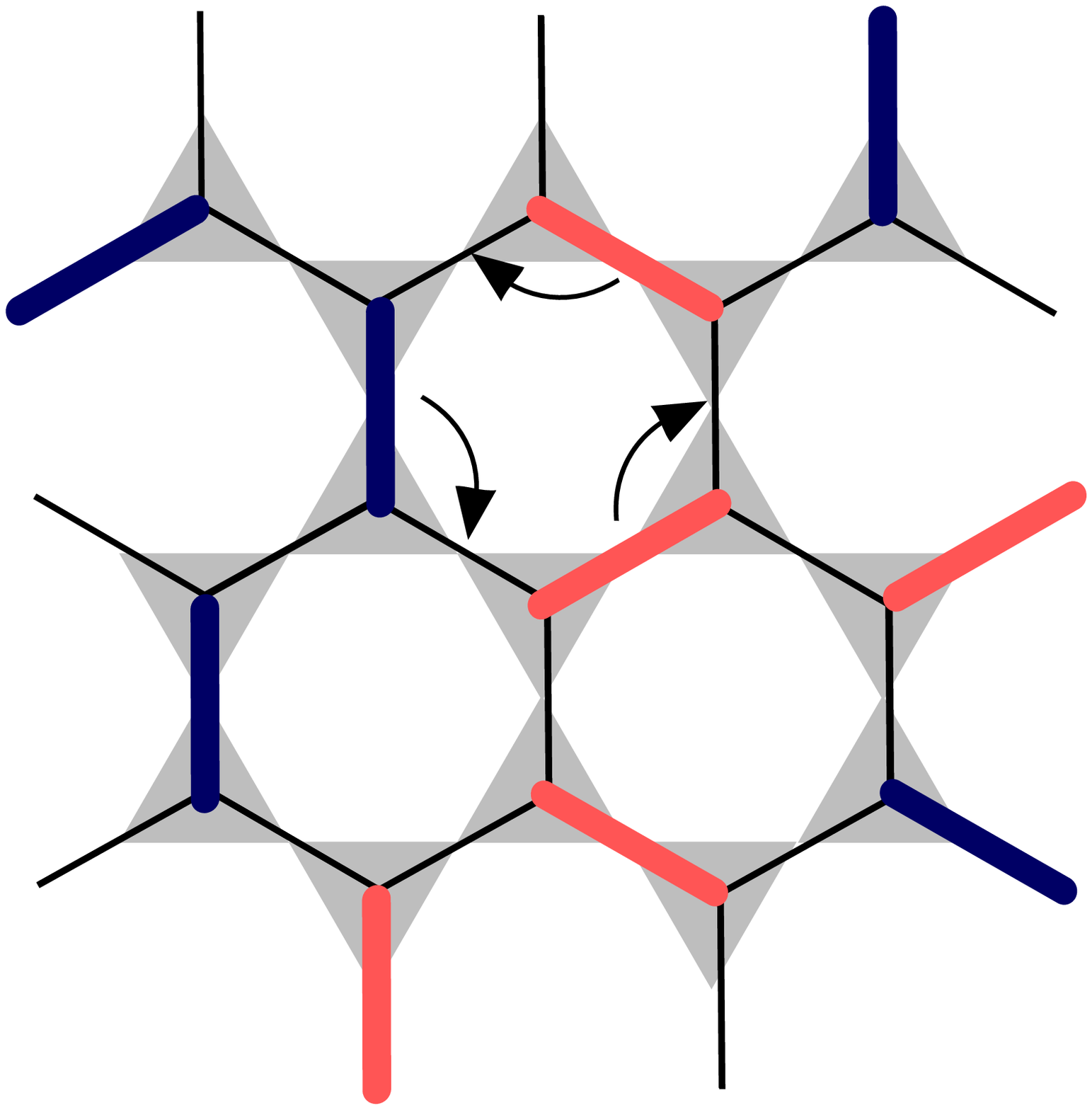}&
(d)\includegraphics[width=28mm]{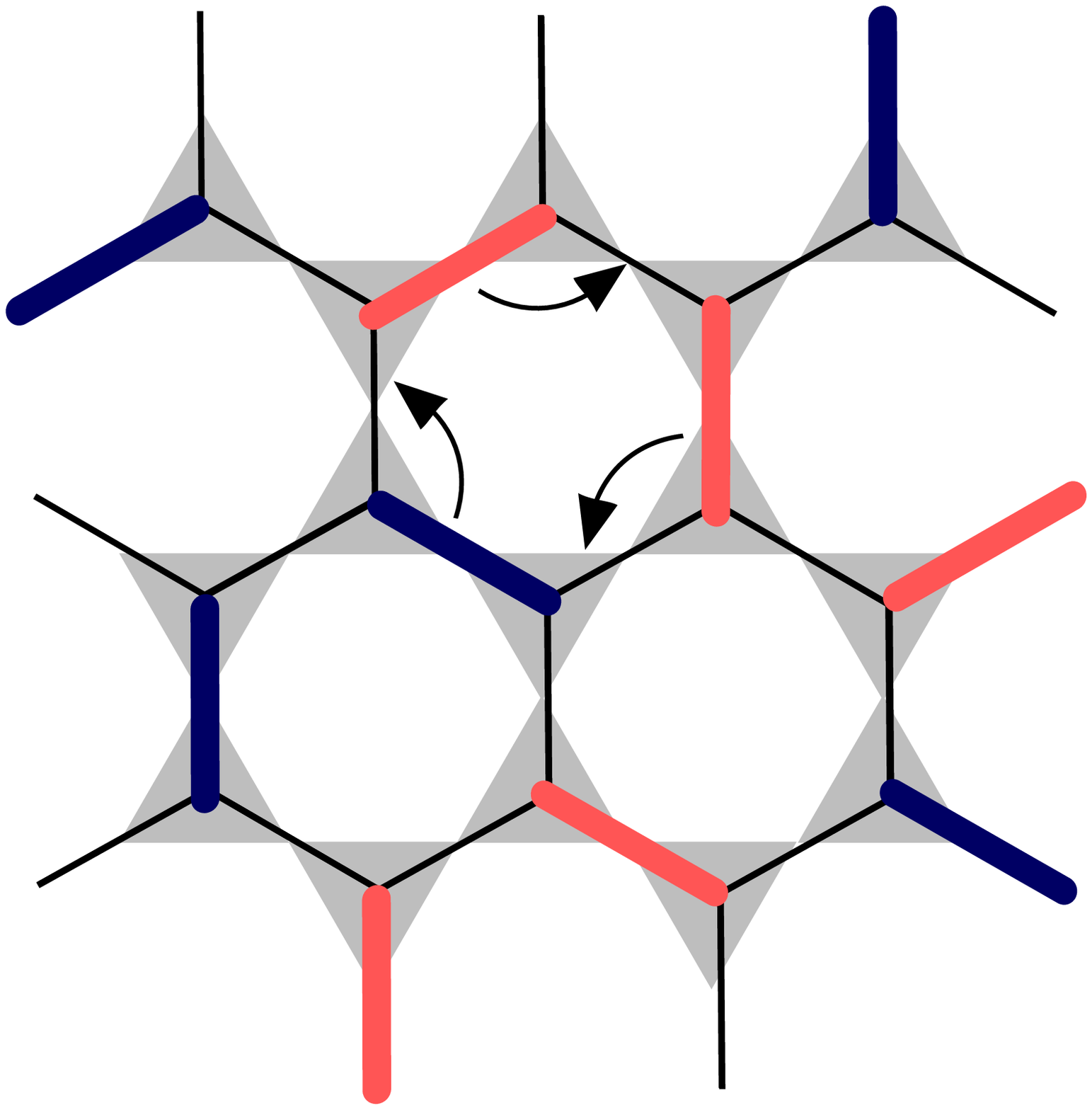}&
\tabularnewline
(e)\includegraphics[width=28mm]{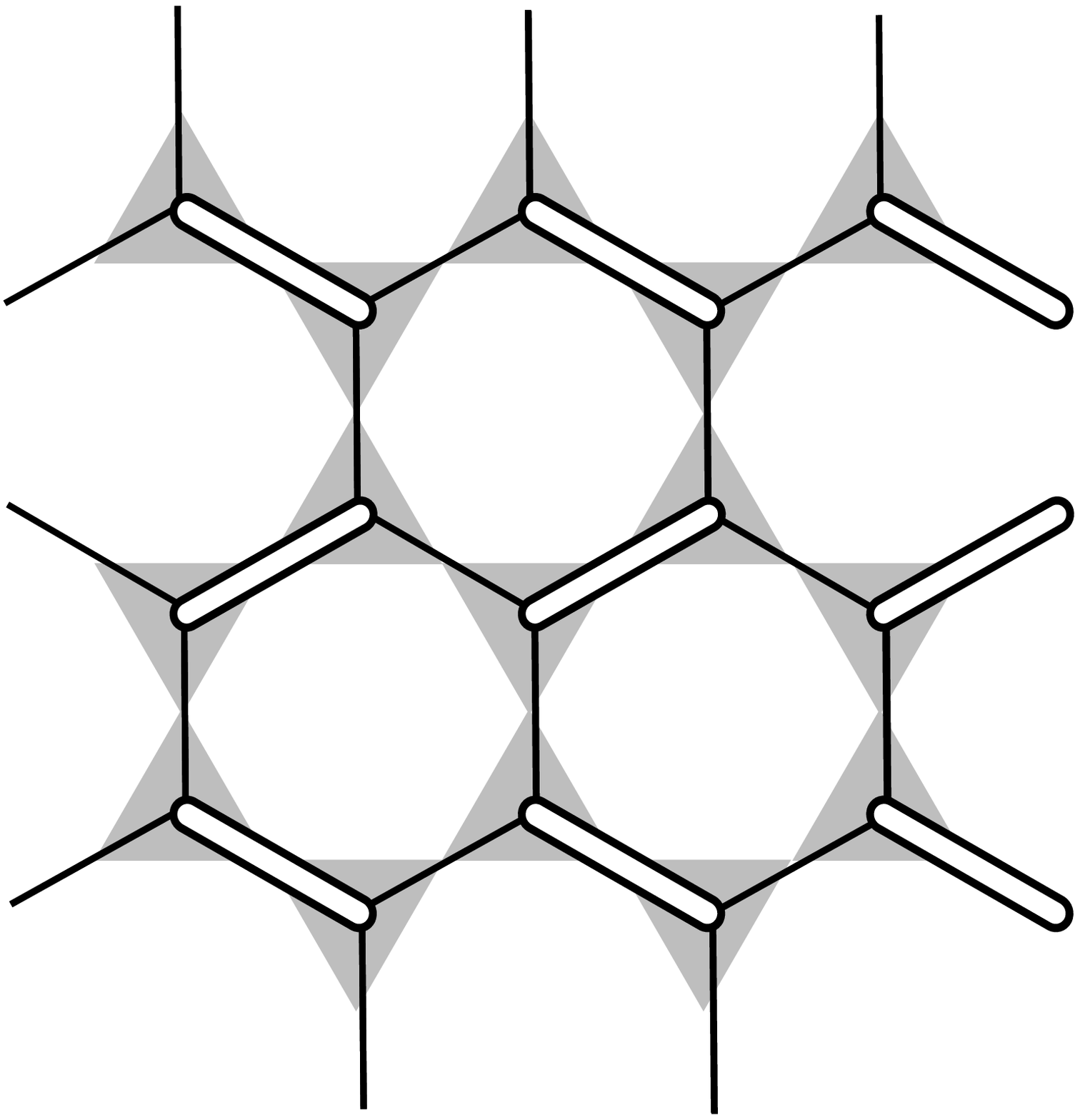}&
(f)\includegraphics[width=28mm]{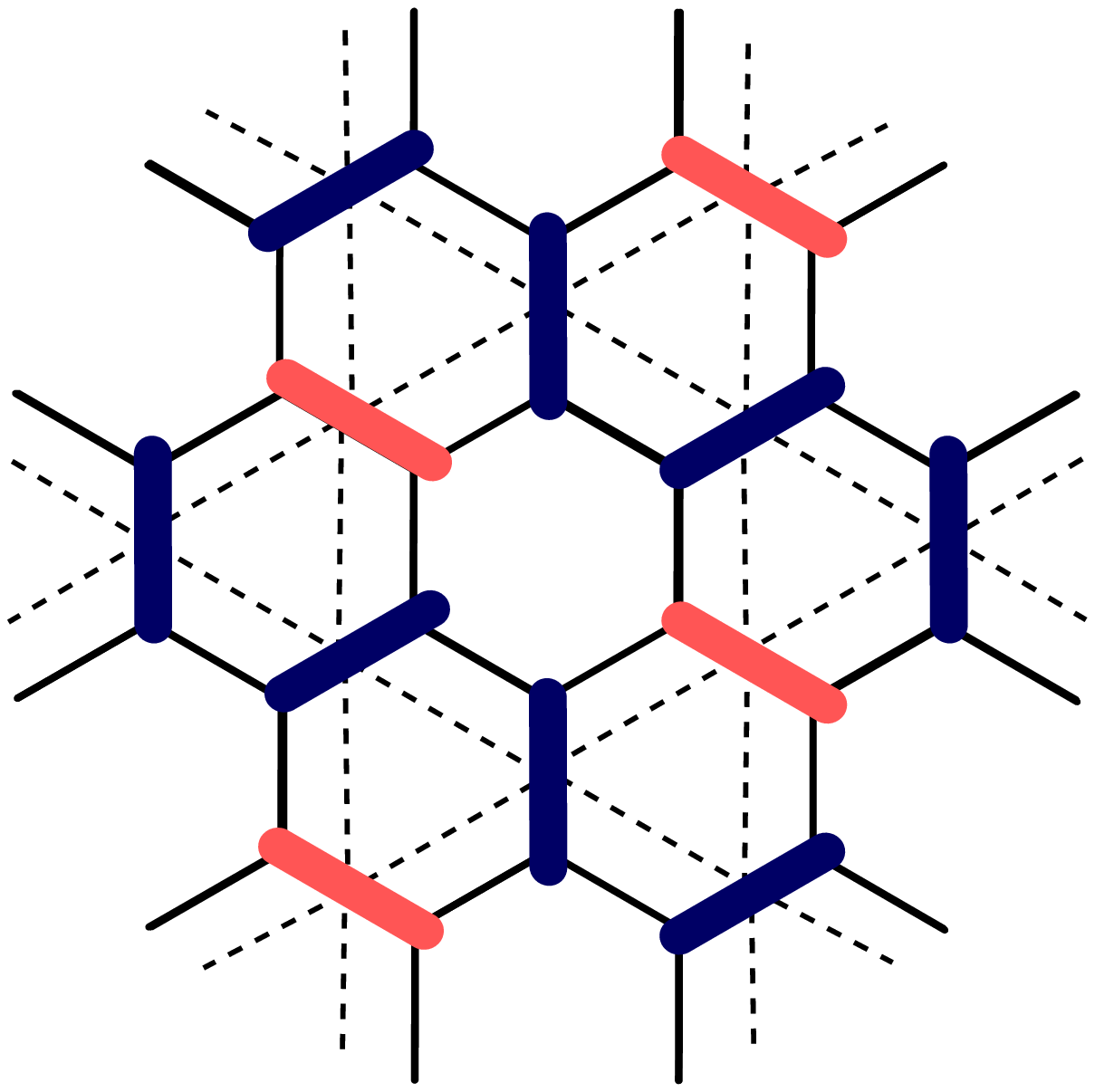}&
\tabularnewline
\end{tabular}\par
\end{centering}
\caption{Panels (a) and (b) show two different configurations
satisfying the constraint of zero or one electron per site and one electron
of arbitrary spin per triangle. The arrows indicate possible ring-hopping
processes. An equivalent colored
dimer representation on a honeycomb lattice is shown in (c) and (d);
the colors encode particle spins. Ring exchanges conserve the parity of the
number of dimers on the sublattice shown in panel (e).
Panel (f) shows the 24-site cluster used for the exact diagonalization. The
dimers are arranged to maximize the next-nearest neighbor spin interaction along
the dashed lines.
\label{dimerpanel}}
\end{figure}

We first focus on the case of 1/6 filling (i.e., one electron per three
sites). In the limit of strong
correlations, when $|t|\ll V\ll U$ and $U\rightarrow\infty$, the possibility
of doubly occupied sites is eliminated. First we assume that $t=0$. In that
case the GS is macroscopically degenerate. All configurations with
precisely one electron of arbitrary spin orientation on each triangle are
GSs (see Figs.~\ref{dimerpanel}(a) and (b)). It is helpful to
consider the honeycomb lattice which connects the centers of triangles of
the kagom\'{e} lattice.
Different GS configurations on the kagom\'{e} lattice correspond
to different two-colored (spin) dimer configurations on the honeycomb
lattice (particles are sitting here on links, see Fig.~\ref{dimerpanel}~(c)
and (d)). They are orthogonal because any wavefunction overlap is neglected.

When $t\ne0$, this GS degeneracy is lifted. In the lowest
non-vanishing order in $t/V$,
the effective Hamiltonian acting within the low-energy manifold spanned by the
states with no double occupancy and exactly one electron per triangle becomes
\begin{equation}
H_{\text{hex}}=-g\sum_{\left\{ \smallhexh\right\}
\left\{\blacktriangle\blacksquare\bullet\right\}}
\Big(\big|\hexaa\big\rangle\big\langle\hexb\big|
+\big|\hexab\big\rangle\big\langle\hexb\big|+\text{H.c.}\Big)
\label{eq:Hhex}
\end{equation}
with $g=6t^{3}/V^{2}$.
Here the Hamiltonian is written in terms of dimers on a honeycomb
lattice  and the sum is performed over all hexagons. The sum over the three
symbols is taken over all
possible color (spin) combinations of a flippable hexagon. Particles
hop either clockwise or counter-clockwise around the hexagons.
These processes can lead to different configurations, depending on the
colors (spins) of the dimers. We observe that $H_{\text{hex}}$ does
not cause a fermionic sign problem. In particular, the local constraint of
having one fermionic dimer attached to each site allows for an enumeration of
dimers such that only an even number of fermionic operators
has to be exchanged when the matrix elements of $H_{\text{hex}}$
are calculated.

If one were to ignore the spin degrees of freedom (the colors of the
dimers), the model would be
equivalent to the quantum dimer model (QDM)
studied in Ref.~\cite{Moessner01c}. Similarly to the
QDM on a square lattice \cite{Rokhsar88}, the effective Hamiltonian
(\ref{eq:Hhex})
conserves certain quantities -- winding numbers -- and connects configurations
only when
they belong to the same topological sector. (For the case of periodic
boundary conditions, the winding numbers are defined by first orienting
dimers so that the arrows point from the A to B sublattice, and second,
by counting the net flow of these arrows across two independent essential cycles
formed by the dual bonds.) The
GS of the QDM was found to be three-fold degenerate in the
thermodynamic limit, corresponding to the valence bond solid (VBS) plaquette
phase with broken translational invariance. In what follows, we
shall investigate the effects of quantum dynamics -- the ring-exchange hopping
of electrons (dimers) -- on spin correlations. Note that $H_{\text{hex}}$
has no explicit spin dependency and conserves both $S^{z}_{\text{tot}}$ and the total spin
$S_{\text{tot}}$.

\paragraph*{Ferromagnetism from the Perron--Frobenius Theorem.}
In short, the Perron--Frobenius theorem states
that the largest eigenvalue of a symmetric $n\times n$ matrix with only
positive elements is  positive and non-degenerate, while the corresponding
eigenvector is ``nodeless'', i.e., can be chosen to have only positive
components. (For a simple proof of this theorem, see e.g.
\cite{Ninio}.)
Applying this theorem to the (finite-dimensional) matrix
$\exp{(-\tau\hat{H})}$ (for any $\tau>0$), one concludes
that if all off-diagonal matrix elements of the Hamiltonian $\hat{H}$ are
non-positive and the Hilbert space is connected by the quantum dynamics
(meaning that any state can be reached from any other state by a repeated
application of $\hat{H}$), then the GS is unique and
nodeless. It is important to remember that the theorem only works for
systems with a finite-dimensional Hilbert space.

To show the relation of this theorem to ferromagnetism, we now sketch the
argument for Nagaoka's ferromagnetism  in the GS of an infinite
$U$ Hubbard model (Eq.~(\ref{eq:extended_hub}) with $V=0$, $U\to \infty$)
with a single mobile hole (after
Refs.~\cite{Tasaki89,Tian90}).  Denote a state with a
single electron or hole on a site $i$ as $\left|i,\alpha\right\rangle$ where
$\alpha=\left\{\sigma_1,\ldots
\sigma_{i-1},\sigma_{i+1},\ldots \sigma_{N}\right\}$ is
the spin configuration of electrons. We use the convention
$\left|i,\alpha\right\rangle= (-1)^i c_{1,\sigma_1}^\dag,\ldots
c_{i-1,\sigma_{i-1}}^\dag
c_{i+1,\sigma_{i+1}}^\dag,\ldots c_{N,\sigma_{N}}^\dag |0\rangle$.
(No double occupancy is allowed if $U\to \infty$.) In
this basis, the matrix elements of the hopping Hamiltonian are either $t$,
for the states related by a single hop of the hole between two neighbor
sites, or 0 otherwise. The Hamiltonian commutes with both
$\hat{S}^z_\text{tot}$ and $\hat{S}^2_\text{tot}$. Our chosen basis consists
of eigenstates of $\hat{S}^z_\text{tot}$ but not $\hat{S}^2_\text{tot}$,
hence we can immediately separate the Hilbert space into sectors of fixed
$S^z_\text{tot}$. Within each sector, the Hamiltonian matrix has exactly $z$
(the coordination number) nonzero entries in each row and each column.
A direct inspection shows that a vector whose entries are all 1 is an
eigenvector with the eigenvalue $zt$. If $t<0$ and tunneling of a
single hole satisfies the connectivity condition, the Perron-Frobenius
theorem applies and hence such a state is the GS (there can be
no other state with positive coefficients only that is orthogonal to this one),
which is unique for a finite system. (We remark that the sign of $t$ can
always be changed on a bipartite lattice.)
Clearly, this state is fully spin-polarized in the
$S^z_\text{tot}=S_\text{max}=\pm N/2$ sectors, and since the Hamiltonian
commutes with $\hat{S}^2_\text{tot}$, the state with
${S}^2_\text{tot}=(S_\text{max}+1)S_\text{max}$ must have the same energy in
every $S^z_\text{tot}$ sector. But we already saw that the states with the
energy $\mathcal{E}=zt$ are unique GSs in every sector, hence they
must have the same ${S}^2_\text{tot}=(S_\text{max}+1)S_\text{max}$, QED.
The obvious pitfalls may come from taking the thermodynamic limit or violating
the connectivity condition: in both cases the nodeless, fully spin polarized
state remain a GS but no claims can be made about other potential GSs.

Turning to our case, we remark that the sign of the plaquette flip in
Eq.~(\ref{eq:Hhex}) can be always chosen negative, irrespective of the sign
of the original tunneling amplitude $t$ -- this is just a matter of a simple
local gauge transformation \cite{Rokhsar88}.
Specifically, the sign of $g$ in Eq.~(\ref{eq:Hhex}) can be changed by
multiplying all configurations $C$ with the color-independent factor
$i^{\nu(C)}$ where $\nu(C)$ is the number of dimers on the sublattice shown in
Fig.~\ref{dimerpanel}~(e). This fact might appear surprising
though since the actual sign of $t$ can typically be gauged away  only for the
cases of bipartite or half-filled lattices.
The reason the sign of $t$ turns out
to be inconsequential in our case of the (non-bipartite) kagom\'{e} lattice
away from half-filling is
due to the constrained nature of the ring exchange quantum dynamics of
Eq.~(\ref{eq:Hhex}). We therefore choose all off-diagonal matrix elements of
$H_{\text{hex}}$ to be non-positive. This by itself is not yet sufficient to
apply the Perron--Frobenius theorem since the quantum dynamics of dimers on a
(bipartite) honeycomb lattice is explicitly non-ergodic: as we have mentioned
earlier, the Hilbert space is broken into sectors corresponding to the winding
numbers which are conserved under \emph{any} local ring exchanges. On the
other hand, the ring-exchange dynamics of dimers given by Eq.~(\ref{eq:Hhex})
\emph{is} ergodic within each sector \cite{Saldanha95}. Therefore we consider
each topological sector separately. The argument is very similar to the one
presented earlier for Nagaoka's ferromagnetism. For the
$S^z_\text{tot}=S_\text{max}=\pm N_\text{e}/2$ spin sectors, the GS
is unique, fully spin-polarized, and all elements of its eigenvector are
positive.

For all other $S^z_\text{tot}$ sectors, however, the situation appears more complicated at first sight. The reason is a much bigger configuration space -- essentially, we are now dealing with two-color dimer configurations. A given state can now be connected by a ring-exchange Hamiltonian to a larger number of states than it would if all dimers had the same color (spin).
We formalize this by introducing the notion of descendant states $|C^{k}_i\rangle,\ k=1\dots 2^{N_\text{e}}$ -- two-color dimer configurations obtained from the uncolored ``parent'' configuration $|C_i\rangle$ by simply coloring its
dimers (i.e., assigning spins). The subspace of descendant states can be partitioned according the conserved $S^z_\text{tot}$. The resulting sectors, in general, have different dimensionality $D(S^z_\text{tot})$ equal to the number of distinct permutations of spins (colors). A crucial observation is that $\sum_{k} \langle C^{k}_i|H_{\text{hex}}|C^{m}_j\rangle = \langle C_i|H_{\text{hex}}|C_j\rangle$ for any $i$, $j$, $m$. The immediate consequence is that if $|\Psi_0\rangle \equiv |\Psi_0(N_\text{e}/2)\rangle = \sum_i \gamma_i |C_i\rangle$ is the GS in the $S^z_\text{tot}=S_\text{max}=\pm N_\text{e}/2$ spin sector, then $|\Psi_0(S^z_\text{tot})\rangle = D^{-1/2}(S^z_\text{tot}) \sum_i \gamma_i \sum_k^\prime |C^{k}_i\rangle$ is an eigenstate with the same energy in any other $S^z_\text{tot}$ spin sector. (The sum over descendants $k$ is performed only within a given spin sector.)
The SU(2) symmetry of the effective Hamiltonian (\ref{eq:Hhex}) once again implies that $|\Psi_0(S^z_\text{tot})\rangle$ is a GS and is fully spin polarized.

Notice that the \emph{uniqueness} of such a GS relies on the ergodicity of the Hamiltonian within each  $S^z_\text{tot}$ spin sector. Numerical studies on finite clusters up to 48 kagom\'{e} lattice sites including different geometries show that this is in fact the case. Unfortunately, we were not able to provide a rigorous analytical argument.  Should it turn \emph{not} to be the case, it would open a possibility for degenerate GSs in  $S^z_\text{tot}\neq S_\text{max}$ spin sectors. Still, at least one of the GSs is fully spin polarized \cite{Tasaki98}.

While the Perron--Frobenius argument applies to finite systems, it does not withstand the thermodynamic limit. In particular, it is known that in the thermodynamic limit the ground state of the QDM is in the three-fold degenerate plaquette phase \cite{Moessner01c}. We suspect that ferromagnetism survives this limit and coexists with such a broken symmetry state; a conclusive resolution of this point remains a subject of further research.

The ferromagnetic GS which we find here should not be mistaken with Mielke's
flat-band ferromagnetism \cite{Mielke91a,Mielke92}. In fact, Mielke has shown that a positive-$U$ Hubbard model with $V=0$ on a kagom\'{e} lattice at 5/6 filling has a fully spin-polarized GS. A detailed discussion of the differences to our case can be found below.

\paragraph*{Stability of kinetic ferromagnetism.}

In order to test the robustness of the ferromagnetic GS,
we introduce  by hand an additional next-nearest neighbor interaction, in the spirit of \cite{Poilblanc07}:
\begin{equation}
H'= H_{\text{hex}}+J\sum_{\langle\langle
i,j\rangle\rangle}\left(S_{i}S_{j}-\frac{1}{4}n_{i}n_{j}\right).\label{
eq:Hspin}\end{equation}
By adding such a term, we attempt to frustrate the ferromagnetic state.  Indeed, this term favors configurations such as the one shown  in Fig.~\ref{dimerpanel}(f) which maximize the spin interactions by having electrons on the same sublattice. Not only the resulting charge order is expected to suppress the kinetic mechanism for ferromagnetism, the ferromagnetic order itself is now suppressed by antiferromagnetic fluctuations favored by the spin interactions for $J>0$.By gradually increasing $J/g$ towards strong antiferromagnetic coupling,
we can estimate the stability of the ferromagnetic GS under the presence of short ranged perturbations.

\paragraph*{Numerical Results.}
\begin{figure}
\begin{centering}\includegraphics[width=87mm]{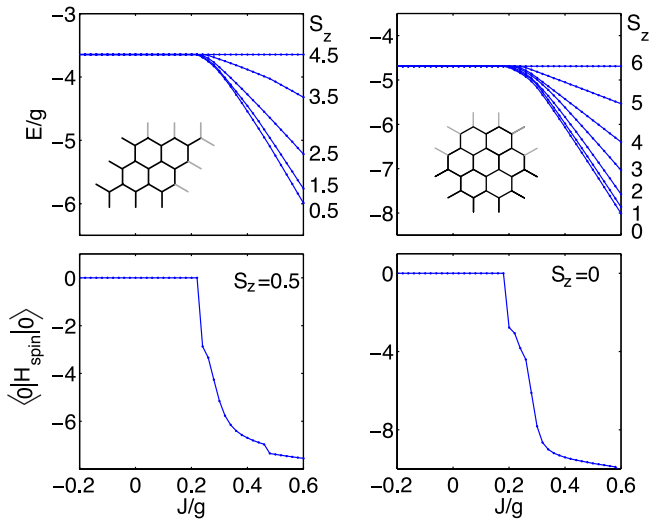}\par\end{centering}
\caption{Exact diagonalization of the two-color dimer model on a 18-site
(left panels) and 24-site (right panels) honeycomb cluster. The upper
panels show the GS energies of different $S_{z}$ sectors
as a function of next-nearest neighbor coupling $J/g$. The lower
ones show the GS expectation values of the spin part of the
Hamiltonian.
\label{fig:Exact-diag} }
\end{figure}

We calculate by means of exact diagonalization the GS of
a two-color dimer model on a 18- and 24-site honeycomb lattice.
The system corresponds to a 1/6 filled kagom\'{e} cluster with 27 and
36 sites, respectively. The calculations on the clusters of different
size and geometry show qualitatively the same results.

The ground-state energies of the different $S_{z}$ sectors are degenerate
as long as $J/g<J/g)_{c}\approx0.2$, as shown in
Fig.~\ref{fig:Exact-diag}.
This demonstrates the robustness of ferromagnetism induced by ring-hopping
processes. Above the transition point $(J/g)_{c}\approx0.2$ antiferromagnetic
spin fluctuations are no longer suppressed. The ground-state degeneracy
of the different $S_{z}$ sectors is lifted and the true GS is
the one with the lowest $|S_{z}|$. The gain in kinetic energy decreases
correspondingly since the spin fluctuations cause nodes in the wavefunction.
The expectation value of $H_{\text{spin}}$ shows a jump at $(J/g)_{c}$
(see Fig. ~\ref{fig:Exact-diag}). Note that a small second jump
at $J/g\approx0.45$ found for the 18 site cluster is an effect of
geometry. The observed findings demonstrate a considerable robustness
of ferromagnetism generated by kinetic processes.
In the limit $J/g\rightarrow\infty$, the kinetic processes are unimportant
and the GS is that of a Heisenberg antiferromagnet on a
kagom\'{e} lattice. One of the configurations which maximize the spin
interactions is shown in Fig.~\ref{dimerpanel}~(f).

The above considerations can also be applied to the case of 5/6 filling due
to the arbitrary choice of sign of $g$ in Eq.~(\ref{eq:Hhex}).
In addition, a ferromagnetic GS is found numerically for a filling
factor of 1/3. With two occupied sites per triangle, we obtain
a fully packed loop covering of the honeycomb lattice instead of a
dimer covering. As before, there is no sign problem in the strong coupling
limit. However,  the effective Hamiltonian is \emph{not} ergodic in  the 1/3
filled case and thus the Perron--Frobenius theorem does not rule out other
GSs which are not fully spin polarized. Numerical
studies confirm a ferromagnetic GS which is much less robust --
antiferromagnetic fluctuations occur already at  very small ratios $(J/g)$. A
more detailed discussion is left to an extended version of this paper.

We conclude by reiterating the difference between the two mechanisms for
ferromagnetism in a Hubbard model on the kagom\'{e} lattice: the one presented
here and the flat-band mechanism discussed in
Refs.~\cite{Mielke91a,Mielke92,Tasaki98}. The flat-band ferromagnetism
was demonstrated for the case of $V=0$ in the Hamiltonian
(\ref{eq:extended_hub}), while our mechanism requires  $V \to
\infty$. In Mielke's case, ferromagnetism has been predicted for the range of
fillings between $5/6$ and $11/12$ and any value of $U>0$. The connection between the sign of the tunneling amplitude $t$ and the electron concentration is crucial. This
is  because the tight binding model on a kagom\'{e} lattice has one of its
three bands completely flat: the lowest band for the case of $t<0$ or the
highest band for the case of $t>0$.
On the other hand, the mechanism presented here
is insensitive to the sign of $t$ and, as we have already mentioned in the
introduction, is of kinetic rather than of potential origin. Furthermore, Perron--Frobenius theorem  is not applicable in Mielke's case \cite{Tasaki98}, instead
the proof was based on graph-theoretical methods. By contrast, the
kinetic ferromagnetism studied in this letter relies crucially on strong
electron--electron repulsion $U \gg V\to \infty$; hence it belongs to a
different class from flat-band ferromagnetism.
Despite certain similarities, it must also be
distinguished from Thouless' three-particle ring exchange mechanism: the
crucial difference is that a standard three-particle ring exchange leaves the
particles at the original locations, it simply cyclically permutes them. In
our case the particles actually move; the initial and final configurations are
distinct.

Thus the mechanism found in this letter represents a new generic type of kinetic ferromagnetism.  An interesting question is to what extent the ferromagnetism of the form found here can be extended to other lattice structures. A particularly interesting case is the pyrochlore lattice. For the strong correlation limit at $1/8$ filling one can use the Perron--Frobenius-based argument as well, implying a ferromagnetic GS. Again, this and other cases will be discussed in the extended version of this paper.

The authors would like to thank G.~Misguich  and R.~Kenyon for illuminating
and helpful discussions.

\end{document}